\magnification=1200

\font\bigbf=cmbx10  scaled\magstep2 \vskip 0.2in
\centerline{\bigbf Aharonov-Casher effect   } \vskip 0.1in
\centerline{\bigbf for spin one particles in a noncommutative
space }

\vskip 0.4in \font\bigtenrm=cmr10 scaled\magstep1
\centerline{\bigtenrm  B. Mirza, R. Narimani and M. Zarei }
\vskip 0.2in

\centerline{\sl Department of Physics, Isfahan University of
Technology, Isfahan 84156, Iran }

\vskip 0.1in

\centerline{\sl E-mail: b.mirza@cc.iut.ac.ir}

\vskip 0.2in \centerline{\bf ABSTRACT} \vskip 0.1in

In this work the Aharonov-Casher (AC) phase is calculated for spin
one particles in a noncommutative space. The AC phase has
previously been calculated from the Dirac equation in a
noncommutative space using a gauge-like technique [17]. In the
spin-one, we use kemmer equation to calculate the phase in a
similar manner. It is shown that the holonomy receives
non-trivial kinematical corrections. By comparing the new result
with the already known spin $1/2$ case, one may conjecture a
generalized formula for the corrections to holonomy for higher
spins.



\vfill\eject

\vskip 1in


\noindent { \bf 1 \   Introduction } \vskip 0.1in

\noindent In the last few years, theories in noncommutative space
have been studied extensively (for a review see [1]).
Noncommutative field theories are related to M-theory
compactification [2], string theory in nontrivial backgrounds [3]
and quantum Hall effect [4]. Inclusion of noncommutativity in
quantum field theory can be achieved in two different ways: via
Moyal $\star$-product on the space of ordinary functions, or
defining the field theory on a coordinate operator space which is
intrinsically noncommutative[1,5]. The equivalence between the
two approaches has been nicely described in [6]. A simple insight
on the role of noncommutativity in field theory can be obtained
by studying the one particle sector, which prompted an interest
in the study of noncommutative quantum mechanics
[7,8,9,10,11,12,13,14]. In these studies some attention was paid
to the Aharonov-Bohm effect [15]. If the noncommutative effects
are important at very high energies, then one could posit a
decoupling theorem that produces the standard quantum field
theory as an effective field theory and that does not remind the
noncommutative effects. However the experience from atomic and
molecular physics strongly suggests that the decoupling is never
complete and that the high energy effects appear in the effective
action as topological remnants. Along these lines, the
Aharonov-Bohm and Aharonov-Casher effects have already been
investigated in a noncommutative space [16,17].  In this work, we
will develop a method to obtain the corrections to the
topological phase of the Aharonov-Casher effect for spin one
particles, where we know that in a commutative space the line
spectrum does not depend on the relativistic nature of the
dipoles. The article is organized as follows; in section 2, we
discuss the Aharonov-Casher effect for spin one particles on a
commutative space. In section 3, the Aharonov-Casher effect in a
noncommutative space is studied and a generalized formula for
holonomy is given.


\vskip 0.2in

\noindent { \bf 2 \ The Aharonov-Casher effect } \vskip 0.1in

\noindent In 1984 Aharonov and Casher (AC) [18] pointed out that
the wave function of a neutral particle with nonzero magnetic
moment $\mu $ develops a topological phase when traveling in a
closed path which encircles an infinitely long filament carrying a
uniform charge density. The AC phase has been measured
experimentally [19]. This phenomenon is similar to the
Aharonov-Bohm (AB) effect. The similarities and the differences
of these two phenomena and possible classical interpretations of
the AC effect have been discussed by several authors [20,21,22].
In Ref. [18], the topological phase of the AC effect was derived
by considering a neutral particle with a nonzero magnetic dipole
moment moving in an electric field produced by an infinitely long
filament carrying a uniform charge density. If the particle
travels over a closed path which includes the filament, a
topological phase will result. This phase is given by

$$ \phi_{AC}= \oint ({ {\vec \mu}} \times {\bf E}) \cdot  d{\bf r}  \eqno(1) $$


\noindent   where $\vec \mu =\mu \vec \sigma $ is the magnetic
dipole moment and $\vec \sigma = (\sigma_1,\sigma_2,\sigma_3)$,
where $ \sigma_i \ (i=1,2,3)$ are the $2 \times 2$ Pauli matrices.
It is possible to arrange that the particle moves in the x-y plane
and travels over a closed path which includes an infinite filament
along z-axis. The electric field in the point ${ \bf r} = x \hat i
+ y \hat j $, where $\hat i$ and $ \hat j $ are unit vectors in
the direction of the positive x and y axes, is given as

$$ {\bf E} = {\lambda \over {2 \pi (x^2 + y^2 )}}(x \hat i
+ y \hat j)  \eqno(2) $$

 \noindent where $\lambda $ is the linear charge density of the
filament and the phase is given by

$$  \phi_{AC}= \mu \sigma_3 \oint ({ \hat{k}} \times {\bf E} ) \cdot  d
 {\bf r} =  \mu \sigma_3 \lambda   \eqno(3) $$

\noindent  where $\hat k $ is a unit vector along z axis. This
phase is purely quantum mechanical and has no classical
interpretation. The appearance of $\sigma_3 $ in the phase
represents the spin degrees of freedom. We see that different
components acquire phases with different signs. This is also one
of the points that distinguishes the AC effect from the AB effect
[23]. In this part, we briefly explain a method for obtaining
Eq.(3). The equation of motion for a neutral spin half particle
with a nonzero magnetic dipole moment moving in a static electric
field $\bf E $ is given by

$$ (i \gamma_{\mu} \partial^{\mu}+ {1\over 2}\mu \sigma_{\alpha \beta}F^{\alpha \beta}-m)
 \psi=0 \eqno(4) $$

\noindent or it can be written as

$$ (i \gamma_{\mu} \partial^{\mu}  -i\mu   {\gamma} \cdot {\bf E} \gamma_{0} -m)  \psi=0
\eqno(5) $$

\noindent where   $ \gamma = (\gamma^1, \gamma^2 , \gamma^3)$ and
$ \gamma-$matrices are defined by

$$ \gamma^0= \left(\matrix{I & 0\cr 0 & -I\cr }\right) \ \ \ \ \ \
 \  \   \ \ \ \ \ \ \
 \gamma^i=\left(\matrix{ 0 & \sigma_i \cr -\sigma_i & 0  \cr
 }\right)\ \ \ \ \
 \eqno(6) $$

 \noindent We define

$$ \psi = \  e^{{\bf a} f} \psi_{0} \eqno(7) $$

\noindent where { \bf a } is the matrix to be determined below, $
f$ is a time independent scalar phase, and $ \psi_0 $ is a
solution of the
 Dirac equation

$$ (i \gamma_{\mu} \partial^{\mu}-m)\psi_{0} =0 \eqno(8) $$

\noindent Writing $\psi_0 $  in terms of $\psi $ and multiplying
 (8) by $ e^{{\bf a} f} $  from the left, we obtain

$$ e^{{\bf a} f} (i \gamma^{\mu} \partial_\mu  -m )e^{-{\bf a} f}\psi = 0 \eqno(9) $$

\noindent

$$ (i  e^{{\bf a} f} \gamma^\mu e^{-{\bf a} f}\partial_\mu
- i e^{{\bf a} f} \gamma^i e^{-{\bf a}  f} {\bf a} \  \partial_i f
-m)\psi=0 \eqno(10) $$

\noindent Comparing Eq.(10) with Eq.(5), we find that {\bf a} and
$f $ must satisfy

$$ {\mu}  {\gamma}\cdot {\bf E} \gamma_0 =   ({\gamma} \cdot
{ \nabla}f) {\bf a} \ \ \ , \ \ \ {\bf a} \gamma_\mu = \gamma_\mu
{\bf a} \eqno(11) $$

\noindent The matrix {\bf a} can be expressed by some linear
combination of the complete set of $4 \times 4 $ matrices $ 1,
\gamma_5, \gamma_\mu, \gamma_\mu \gamma_5 $ and $\sigma_{\mu
\nu}={i\over 2}[\gamma_\mu , \gamma_\nu]$. The second member of
Eqs.(11) cannot be satisfied if all $\gamma_1, \gamma_2 $ and
$\gamma_3 $ are present in Eq.(10). However, it is possible to
satisfy it if the problem in question can be reduced to the
planar one. This indicates that the AC topological phase can arise
 only in two spatial dimensions. Therefore, let us consider the particle moving
 in $x-y$ plane in which case only the matrices $\gamma_1 $ and
 $\gamma_2 $ are present in (11), and moreover, $\partial_3 \psi $  and $ E_3 $
 vanish. The choice $-i \sigma_{12}\gamma_0 $ represents a
 consistent Ansatz.  From the first equation in
 (11), we get

$$  \nabla f= \mu ( \hat k \times \bf E) \eqno(12) $$

\noindent and the phase is given by

$$ \eqalignno{\phi^{(0)} & = \sigma_{1 2} \gamma_0  \oint \vec \nabla f\cdot d{\bf r}  \cr
&=\mu \sigma_{1 2} \gamma_0 \oint (\hat{k} \times \bf E )\cdot d
{\bf r} \cr
 &=\mu \left(\matrix{\sigma_3&0\cr
                              0& -\sigma_3\cr}\right)\oint (\hat{k} \times {\bf E} )\cdot d
\bf r & (13) \cr} $$

\noindent One may extend this method to spin one using the first
order kemmer equation (for a more complete explanation and
derivation see [24]). The kemmer equation is defined by

$$ (i \beta^{\mu} \partial_{\mu} -m)\psi =0 \eqno(14) $$

\noindent where the $\beta$-matrices are generalizations of the
Dirac gamma matrices. These satisfy an algebra ring, which for
spin one is

$$ \beta^{\mu} \beta^{\nu} \beta^{\rho}+
\beta^{\rho} \beta^{\nu} \beta^{\mu}= \eta^{\mu \nu
}\beta^{\rho}+\eta^{\nu \rho}\beta^{\mu}  \eqno(15) $$

\noindent These Kemmer $\beta$-matrices are reducible, that is
the $16\times 16$ representation decomposes into three separate
representations: a one dimensional trivial representation; a five
dimensional spin zero representation; and the 10-d spin one
representation [24,25,26].  This algebra is odd, that is it cannot
reduce the matrix operator to the identity, unlike the Dirac
algebra. In this paper we choose the 10-d spin one representation
which is given by the  following $10 \times 10$ matrices

$$ \beta^0 =\left(\matrix{\widehat  O & \widehat O & I &  o^{\dag}\cr
                            \widehat O &\widehat O  & \widehat O & o^{\dag} \cr
                               I &\widehat O  & \widehat O & o^{\dag} \cr
                              o&o&o&0\cr} \right)  \ \ \ \ \ \ \ \
 \beta^j=\left(\matrix{\widehat  O & \widehat  O  & \widehat  O  & -i {K}^{j \dag}\cr
                                 \widehat O & \widehat O  &S^j &  o^{\dag} \cr
                                 \widehat O &-S^j & \widehat O & o^{\dag} \cr
                                 -iK^j &o&o&0\cr}  \right)                \eqno(16)  $$

\noindent where the elements are

$$ \widehat O = \left(\matrix{0 & 0  & 0  \cr
                                 0& 0  & 0  \cr
                                 0 & 0 & 0 \cr}  \right) \ \ \ \ \ \ \ \
                                I = \left(\matrix{1 & 0  & 0  \cr
                                                            0& 1  & 0  \cr
                                                            0 & 0 & 1 \cr}  \right)\eqno(17)$$

$$S^1 = i  \left(\matrix{0 & 0  & 0  \cr
                                 0& 0  & -1  \cr
                                 0 & 1 & 0 \cr}  \right) \ \ \
                                 S^2= i  \left(\matrix{0 & 0  & 1  \cr
                                 0& 0  & 0 \cr
                                 -1 & 0 & 0 \cr}  \right)\ \ \
                                 S^2= i  \left(\matrix{0 & -1  & 0  \cr
                                 1& 0  & 0 \cr
                                 0 & 0 & 0 \cr}  \right)\eqno(18)$$

$$ o = \left(\matrix{0 & 0  & 0 \cr} \right), \ \ \ K^1 = \left(\matrix{1 & 0  & 0 \cr} \right),  \ \ \
   K^2 = \left(\matrix{0 & 1  & 0 \cr} \right),  \ \ \
   K^3 = \left(\matrix{0 & 0  & 1 \cr} \right)  \eqno(19)  $$

\noindent Just as in Dirac theory, the Lorentz invariance of the
Kemmer theory entails a transformation of the spinor so that the
matrix representation remain the same.  The Lorentz generator for
these transformations, $S_{\mu \nu}$, is proportional to the
antysymmetric product of two matrices of the ring,

$$ S_{\mu \nu } =b(\beta_{\mu} \beta_{\nu}-\beta_{\nu}\beta_{\mu})  \eqno(20) $$

\noindent These generators satisfy well known commutation
relation and define the spin operators. The coefficient b is
linked to the coefficient of the commutation relations, and is
set below according to our convenience. The equation of motion
for a neutral spin one particle with an anomalous magnetic moment
in Kemmer theory is

$$ (i\beta_{\mu}\partial^{\mu}+{1\over 2}\mu S_{\alpha \beta}F^{\alpha \beta}-m) \psi=0  \eqno(21) $$

\noindent The interaction term emerge from the derivation of a
second order Kemmer equation following the method of Umezawa [26].
The operator component of the phase in the spin one AC phase
solution is a spin one pseudo-vector operator defined by

$$ \xi_{\mu} = {i\over 2} \varepsilon_{\mu \nu \lambda \rho
}\beta^{\nu} \beta^{\lambda} \beta^{\rho } \eqno(22) $$

The path dependent phase proportional to $\xi_3$ is introduced in
the free Kemmer equation of motion [17]

$$ (i \beta^{\mu} \partial_{\mu}-m )e^{i \xi_3 \int^r {\bf A}
\cdot d{\bf r} } \psi =0 \eqno(23)  $$

\noindent with the intention to transform this into the equation
of motion (21) with the anomalous magnetic moment term.
Multiplying (23) by $e^{(- i \xi_3 \int^r {\bf A} \cdot d{\bf r}
)}$ from the left and comparing with (21), we obtain

$$ exp[{-i \xi_3 \int^r {{\bf A} \cdot d{\bf r} }}]\  \beta^{\nu}  exp[{i \xi_3 \int^r {{\bf A} \cdot d{\bf r}
}}] =  \beta^{\nu}   \eqno(24) $$

$$ -\beta^{\nu} \xi_3 A_{\nu} \psi = {1\over 2} \mu S_{\alpha \beta}F^{\alpha \beta} \psi
=\mu S_{0 i}F^{0 i}\psi \eqno(25)   $$

\noindent By using the Baker-Hausdorff formula in the first
condition, it is easy to see that for $\nu \neq 3$ the commutators
are zero. However for $\nu=3$ the commutators does not vanish and
so the first condition restrict the dynamics to 2+1 dimensions,
just as the spin half.

\noindent By a direct calculation in the second line and using
definition of the $\xi_3$, $\beta^\nu$ and $S_{\mu \nu}$ matrices
one has

$$   A_1= -2 \mu E_2  \ \ \ \ \  \ \ \  A_2= 2 \mu E_1  \eqno(26) $$

\noindent Finally the AC phase for a closed path is given by

$$ \phi_{AC}=\xi_3 \oint {\bf A} \cdot  d{\bf r} =
 2 \mu \xi_3 \int_S (\nabla\cdot E )\cdot dS = 2 \mu \xi_3 \lambda  \eqno(27)$$

\vskip 0.2in

\noindent { \bf 3  \    The spin one AC effect on a
noncommutative space} \vskip 0.1in

\noindent  The noncommutative Moyal spaces can
 be realized as spaces where coordinate operator $ \hat x^\mu $
 satisfies the commutation relations

$$ [ \hat x^\mu , \hat x^\nu ] = i \theta^{\mu \nu } \eqno(28) $$

\noindent where $ \theta^{\mu \nu} $ is an antisymmetric tensor
and is of space dimension (length$)^2 $. We note that space-time
noncommutativity, $\theta^{0 i}\neq 0 $, may lead to some
problems with unitarity and causality. Such problems do not occur
 for the quantum mechanics on a noncommutative space with a usual
 commutative time coordinate. The noncommutative models specified
 by Eq.(14) can be realized in terms of a $\star$-product: the
 commutative algebra of functions with the usual product
 f(x)g(x) is replaced by the $\star$-product Moyal algebra:

 $$( f\star g)(x)= exp \ [{i\over 2} \theta_{\mu \nu
 }\partial_{x_\mu}\partial_{y_{\nu}}]\ f(x)g(y)|_{x=y} \eqno(29)$$

\noindent  As for the phase space, inferred from string theory,
we choose

$$ [\hat x_i , \hat x_j ]= i \theta_{i j}, \ \ \  [\hat x_i, \hat
p_j ]=i\hbar \delta_{ij}, \ \ \ [\hat p_i, \hat p_j]=0. \eqno(30)
$$

\noindent The noncommutative quantum mechanics can be defined by
[7-14],

$$ H( p, x) \star \psi( x)=E \psi( x)\eqno(31) $$

\noindent  The equation of motion for a neutral spin one particle
with a nonzero magnetic dipole moment moving in a static electric
field $\bf E $ is given by

$$ (i \beta_{\mu} \partial^{\mu}+ {1\over 2}\mu S_{\alpha \beta}F^{\alpha \beta}-m)\star \psi=0
\eqno(32) $$

\noindent As in [17] we define

$$ \psi =   e^{i{\xi_3} f} \psi_{0}  \eqno(33) $$

\noindent  where  ${\xi_3}$ is the matrix already defined, $f$ is
a time independent scalar phase, and $ \psi_0 $ is a solution of
the free Kemmer equation

$$ (i \beta_{\mu} \partial^{\mu}-m)\psi_{0} =0 \eqno(34) $$

\noindent By using Baker-Cample-Housdorff formula and
$[\beta_\mu, \xi_3]=0$, (32) can be written as

$$  -\beta^{\mu} \partial_{\mu}(\xi_3 f)   e^{i {\xi_3} f} \psi_{0}+{\mu\over2}S_{\alpha \beta}
F^{\alpha \beta}\star  (e^{i {\xi_3} f} \psi_{0})=0  \eqno(35) $$

\noindent After expanding the second term in (35) up to the first
order of the noncommutativity parameter $ \theta_{ij}= \theta
\epsilon_{ij} $  and defining $k_j $ as

$$ \partial_{j} \psi_{0} = (i k_j ) \psi_{0}  \eqno(36) $$

\noindent the final result up to first order in $\theta $  is
given by

$$ [-\beta^\mu \partial_\mu(\xi_3 f)+ \mu (S_{0l}F^{0 l}+{i\over 2}\theta_{i j}\partial_i
(S_{0l} F^{0l})\partial_j (i \xi_3 f) + {i\over 2}\theta_{i
j}\partial_i (S_{0l} F^{0l})(ik^j))] exp[i \xi_3 f ] \psi_0=0
\eqno(37)
$$

\noindent It should be noted that expansion of $F^{0l}$  or $\bf
E $ up to first order in $\theta $ leads to an additive correction
to the commutative holonomy and does not cause a new
non-topological behaviour. A similar situation happens in the
noncommutative Aharonov-Bohm  effect. By expanding $ f $ up to
first order in $ \theta $

$$ f= f^{(0)}+\theta f^{(1)}+ ... \eqno(38) $$

\noindent we obtain the following equations,

$$ [-\beta^\mu \partial_\mu (\xi_3 f^{(0)}) +\mu (S_{0l}F^{0 l})] \psi_0=0  \eqno(39) $$

 $$ [-\beta^\mu \partial_\mu(\xi_3 f^{(1)})+ {i \mu\over 2}\  \varepsilon_{ij}
   \partial_i(S_{0l} F^{0l}) \partial_j(i\xi_3f^{(0)})
 + {i \mu \over 2} \ \varepsilon_{ij}\partial_i (S_{0l}F^{0l})(ik^j)] \psi_0=0   \eqno(40) $$

\noindent by choosing $b=2$ in $(20)$ and after a straightforward
calculation we get to

$$ \nabla f^{(0)}= 2 \mu (\hat k \times \vec E )  \eqno(41) $$

\noindent which is equivalent to (26), and the phase is given by

$$ \eqalignno{\phi^{(0)} & = \xi_3 \oint {\nabla f^{(0)}} \cdot  d{\bf r}\cr
&= 2 \mu \xi_3 \oint (\hat k \times {\bf E} )\cdot  d{\bf r}& (42)
\cr} $$

\noindent Substituting Eq.(42) in (40) and then using the wave
functions which are given in [24], and a long but straightforward
calculation (the Mathematica package is used) the following
correction to $\phi^{(0)}$ for a neutral particle with nonzero
magnetic dipole moment $ \mu $ and with spin one $(m_s=1,0,-1)$
is obtained

$$ \eqalignno{ \triangle \phi_{\theta} & = \theta \xi_3  \oint \vec \nabla f^{(1)}
\cdot d\vec r \cr &= {\theta \over 2}\xi_3 \varepsilon^{ij} \
\biggl( \mu \oint  k_j (\partial_i E_2 dx_1-
\partial_i E_1 dx_2)   \cr
&  - 2 m_s \oint[(\mu \partial_i  E_2) \ \mu (\hat k \times \vec
E)_j dx_1 - (\mu \partial_i E_1) \ \mu (\hat k \times \vec E)_j
dx_2] \biggr) & (43) \cr}  $$

\noindent where $m_s=1,0,-1$. The first term is a velocity
dependent correction and does not have the topological properties
of the commutative AC effect and could modify the phase shift. The
second term is a correction to the vortex and does not contribute
to the line spectrum. Using $m_s = 1/2,-1/2$ the integral in
 (43) can be mapped to the corrections which have already been
obtained for the spin half Aharonov-Casher effect in [17,
Eq.(32)]. One may conjecture that (43) is also valid for higher
spins. It is interesting to extend these results to higher order
terms, however it seems that obtaining an exact result similar to
the commutative case is not possible by these methods. For some
other interesting relevant papers see [27-34].



\vskip 0.2in
\centerline{\bf \ \  References} \vskip 0.1in

\noindent [1] \ M. R. Douglas and N. A. Nekrasov, Rev. Mod. Phys.
{\bf 73} (2001) 977-1029. hep-th/0106048.

\noindent [2] \ A. Connes, M. R. Douglas and A. Schwarz, JHEP {\bf
9802} (1998) 003, hep-th/9808042.

\noindent [3] \ N. Seiberg and E. Witten, JHEP {\bf 9909} (1999)
032. hep-th/9908142.

\noindent [4] \ L. Susskind, hep-th/0101029.

\noindent [5] \ M. Chaichian, A. Demichev and P. Presnajder, Nucl.
Phys. {\bf B567} 360 (2000), hep-th/9812180.

\noindent [6] \ L. Alvarez-Gaume, S.R. Wadia, Phys. Lett. {\bf
B501} 319 (2001), hep-th/0006219.

\noindent [7] \ L. Mezincescu, hep-th/0007046.

\noindent [8] \ V. P.Nair, Phys. Lett. {\bf B505 } (2001) 249,
hep-th/0008027.

\noindent [9] \ M. Chaichian, M.M. Sheikh-Jabbari and A.Tureanu
Phys. Rev. Lett. {\bf 86}(2001) 2716, hep-th/0010175.

\noindent [10] \ J. Gamboa, M. Loewe and J. C. Rojas, Phys. Rev.
{\bf D64} (2001) 067901, hep-th/0010220.

\noindent [11] \ S. Belluchi , A. Nersessian and C. Sochichiu,
Phys. Lett. {\bf B522} (2001) 345, hep-th/0106138.

\noindent [12] \ C. Acatrinei, JHEP {\bf 0109} (2001) 007,
hep-th/0107078.

\noindent [13] \ J. Gamboa, M. Loewe, F.M. Mendez,J.C. Rojas,
Int. J. Mod. Phys. {\bf A17} (2002) 2555-2566, hep-th/0106125.

\noindent [14] \ J. Gamboa, M. Loewe, F. Mendez and J.C.Rojas,
Mod. Phys. Lett. {\bf A16} (2001) 2075-2078, hep-th/0104224.

\noindent [15] \ Y. Aharonov and D. Bohm, Phys. Rev. {\bf 115}
(1958) 485.

\noindent [16] \ M. Chaichian, A. Demichev, P. Presnajder, M.M.
Sheikh-Jabbari and A. Tureanu, Phys. Lett. {\bf B527}, 149-154
(2002), hep-th/0012175.

\noindent [17] \ B. Mirza, M. Zarei, Eur. Phys. J. C. {\bf
32},583-586 (2004), hep-th/0311222.

\noindent [18] \ Y. Aharonov and A. Casher, Phys. Rev. Lett. {\bf
53} (1984) 319.

\noindent [19] \ A. Cimmino, G.I. Opat, A.G. Klein, H.Kaiser, S.A.
Werner, M.Arif and R. Clothier, Phys. Rev. Lett. {\bf 63} (1989)
380.

\noindent [20] \ A.S. Goldhaber, Phys. Rev. Lett. {\bf 62} (1989)
482.

\noindent [21] \ T.H. Boyer, Phys. Rev. {\bf A36} (1987) 5083.

\noindent [22] \ Y. Aharonov, P. Pearl and L. Vaidman, Phys. Rev.
{\bf A37} (1988) 4052.

\noindent [23] \ Xiao-Gang He and Bruce H.J. McKeller, Phys Lett.
{\bf B256} (1991) 250.

\noindent [24] \ J. A. Swansson and B. H. J. McKellar, J. Phys. A:
Math Gen. {\bf 34} (2001)  1051-1061, quant-ph/0007118.

\noindent [25] \ W. Greiner, Relativistic Quantum Mechanics,
Springer-Verlag, 1990.

\noindent [26] \ H. Umezawa, Quantum Field Theory, North-Holland,
1956.

\noindent [27] \ I. Hinchliffe, N. Kersting, Y.L. Ma, Review of
the Phenomenology of Noncommutative Geometry,Int.J.Mod.Phys. A19
(2004) 179-204, hep-ph/0205040.

\noindent [28] \ S. A. Alavi, Mod. Phys. Lett. {\bf A20} (2005)
1013.

\noindent [29] \ Li, Kang; Dulat, Sayipjamal, Eur. Phys. J. {\bf
C46} (2006) 825-828.

\noindent [30] \ B. Basu, S. Ghosh, S. Dhar, hep-th/0604068.

\noindent [31]\ P. Schupp, J. Trampetic, J. Wess and G. Raffelt,
Eur. Phys. J. {\bf C36} (2004) 405-410, hep-ph/0212292.

\noindent [32] \ Ya. I. Azimov and R. M. Ryndin, Phys. Atom. Nucl.
61 (1998) 1932-1936, hep-ph/9710433.

\noindent [33] \ C. Duval and P. A. Horvathy, J. Phys. {\bf A34}
(2001) 10097-10108, hep-th/0106089.

\noindent [34] \  M. Haghighat, M. M. Ettefaghi, M. Zeinali,
Phys.Rev. {\bf D73} (2006) 013007

\noindent [35] \  L.P. Colatto , A.L.A. Penna , W.C. Santos,
Phys.Rev. {\bf D73} (2006) 105007.

\noindent [36] \  E. Passos, L. R. Ribeiro, C. Furtado, J. R.
Nascimento, hep-th/0610222.

\vfill\eject

\bye